\documentclass[authoryear,square,12pt,3p]{jowarticle}

\usepackage[english]{babel}
\usepackage[utf8]{inputenc}
\usepackage{amsmath}
\usepackage{graphicx}
\usepackage[colorinlistoftodos]{todonotes}
\usepackage{natbib}

\makeatletter
\renewcommand\section{\@startsection{section}{1}{\z@}{-3.25ex plus -1ex minus -.2ex}{1.5ex plus .2ex}{\normalsize\bf}}
\renewcommand\subsection{\@startsection{subsection}{2}{\z@}{-3.25ex plus -1ex minus -.2ex}{1.5ex plus .2ex}{\normalsize\bf}}
\renewcommand\subsubsection{\@startsection{subsubsection}{3}{\z@}{-3.25ex plus -1ex minus -.2ex}{1.5ex plus .2ex}{\normalsize\bf}}
\makeatother

\begin{document}
\begin{frontmatter}
\title{How to Beat Science and Influence People:\\ Policy Makers and Propaganda in Epistemic Networks}

\author{James Owen Weatherall}\ead{weatherj@uci.edu} \author{Cailin O'Connor}\ead{cailino@uci.edu}
\address{Department of Logic and Philosophy of Science \\ University of California, Irvine}

\author{Justin P. Bruner}\ead{justinpbruner@gmail.com}
\address{Department of Theoretical Philosophy \\ University of Groningen}

\date{ }

\begin{abstract}
In their recent book \emph{Merchants of Doubt} [New York:Bloomsbury 2010], Naomi Oreskes and Erik Conway describe the ``tobacco strategy'', which was used by the tobacco industry to influence policy makers regarding the health risks of tobacco products.  The strategy involved two parts, consisting of (1) promoting and sharing independent research supporting the industry's preferred position and (2) funding additional research, but selectively publishing the results.  We introduce a model of the Tobacco Strategy, and use it to argue that both prongs of the strategy can be extremely effective---even when policy makers rationally update on all evidence available to them.  As we elaborate, this model helps illustrate the conditions under which the Tobacco Strategy is particularly successful.  In addition, we show how journalists engaged in `fair' reporting can inadvertently mimic the effects of industry on public belief.  \end{abstract}
\end{frontmatter}

\section{Introduction}

In December 1952, \emph{Readers Digest}---the publication with the largest circulation in the world at the time---published an article with the title ``Cancer by the Carton'' (\citet{norr1952cancer}).\footnote{The history presented in these opening paragraphs draws heavily on the account given by \citet[Ch. 1]{Oreskes+Conway}.  Note that the chronology we describe differs slightly from theirs, as they report that the \emph{Reader's Digest} article was published in the wake of the 1953 Wynder article, whereas in fact it appeared almost a year earlier.  See also the discussion in \citet{OConnor+Weatherall}.}  Coming on the heels of a period during which smoking rates had risen significantly, the article brought the growing body of evidence that cigarette smoking caused cancer to the attention of a wide audience.  Six months later, a group of doctors at Sloan-Kettering Memorial Hospital completed a study in which they demonstrated that mice painted with cigarette tar developed malignant carcinomas (\citet{wynder1953experimental}).  To a public already primed by the \emph{Reader's Digest} article, this paper provided a direct and visceral link between a known by-product of smoking and fatal cancer.  It produced a media frenzy, with articles reporting the result in a number of national newspapers and magazines.

The tobacco industry panicked.  The heads of six major U.S. tobacco companies contracted with the New York public relations firm Hill and Knowlton to combat claims that their product had serious adverse health effects.  At a meeting in December of 1953, they agreed on a novel strategy: they would fight science with science.  In 1954 they hired a prominent geneticist, C. C. Little, to run the newly create Tobacco Industry Research Committee (TIRC).  The goal of this committee was ostensibly to support and promote research on the health effects of tobacco.  But in fact, it was a propaganda machine whose principal activity was to find, fund, and promote scientific research that contradicted the growing consensus that smoking kills.  The goal was to create a false sense of uncertainty and controversy, so as to delay policy change in the service of public health.

As historians \citet{Oreskes+Conway} painstakingly document, drawing on documents released as part of a series of lawsuits in the 1990s, the tobacco industry initiated this strategy---which Oreskes and Conway dub the ``Tobacco Strategy''---despite the fact that their own internal research had already established a link between smoking and various health risks.\footnote{For other discussions of this strategy, see for instance \citet{michaels2005manufacturing,MichaelsDISP}.}  The Tobacco Strategy had multiple parts.  One part, which we will call \emph{selective sharing}, consisted of identifying and promoting research produced independently of industry that happened to support the industry's preferred position.  For instance, in 1954 the TIRC distributed a pamphlet entitled ``A Scientific Perspective on the Cigarette Controversy'' to nearly 200,000 doctors, journalists, and policy-makers, in which they emphasized favorable research and questioned results supporting the contrary view.  A second part of the Tobacco Strategy, which we will call \emph{biased production}, consisted of funding their own research, either through in-house scientists or by offering research grants to scientists pursuing research activities that the industry supported, and then publishing the results of that research only if it supported industry views.

\citet{Oreskes+Conway} make a compelling case both that the Tobacco Strategy was self-consciously adopted by industry to influence public opinion, and that those goals were apparently achieved.  They also argue that the Tobacco Strategy succeeded, in the sense of playing a causal role in realizing these ends.  But the historical analysis they provide is necessarily \emph{in situ}, and the details are complex.  It is difficult to clearly distinguish the causal role played by the TIRC's activities from other cultural and psychological considerations that may have influenced public opinion and slowed policy-change, such as the facts that smoking was culturally ingrained, including among doctors and legislators, and that the tobacco industry made significant political contributions.  One might worry that it was over-determined that cultural and political change regarding tobacco would be slow, so that it is not clear that the Tobacco Strategy played a significant role.

In this paper, we attempt to strengthen the causal link between the Tobacco Strategy and public opinion by proposing and analyzing an agent-based model of the Tobacco Strategy.  Our model is based on the network epistemology framework introduced by \citet{venkatesh1998learning}, later adapted to philosophy of science by \citet{zollman2007communication}, and used more recently by \citet{holman2015problem,holman2017experimentation} to study industry effects on science.\footnote{Similar methods are used by \citet{OConnor+Weatherall}, \citet{OConnor+WeatherallPolarization}, and \citet{OConnor+WeatherallConformity} to discuss the propagation of false beliefs in social networks more generally.}  This approach allows us to test counterfactual effects of industry propaganda on public belief, free from compounding factors.\footnote{\label{causal} To be clear, we take this ability to intervene in our model to be the key aspect that helps us strengthen the causal link between the Tobacco Strategy and public policy related to tobacco products.  The model allows us to test where and how the presence of a propagandist can influence public belief.  It also lets us explore the effects of a propagandist in isolation from other potential causal factors.  That said, we do not claim that the factors we consider here are the \emph{only} relevant causal factors, nor that other models could not be used to identify other counterfactual relationships and possible causal factors.  For more on the interventionist account of causation that is lurking in the background here, see \citet{woodward2005making}; see also note \ref{howpossibly} for more on the relationship between the model we discuss and ``how-possibly'' modeling.}  As we will argue, our model provides evidence that the Tobacco Strategy can, in principle, be very effective.  We also provide important insight into how (and when) it works.  As we will argue, the Tobacco Strategy trades on the fact that empirical research into difficult, probabilistic problems will yield results that, individually, support more than one position.  It is only by viewing a maximal, unbiased sample of the output of a research community that one can hope to come to hold reliable beliefs about the problem at hand.  This means that even subtle manipulations, not in the scientific record itself, but in the sample of that record that one is exposed to, can have disproportionate impact on one's beliefs\footnote{This conclusion is in line with prior work in social epistemology exploring how subjects come to form incorrect beliefs on the basis of selective reporting (for an overview, see \citet{romero2016can}).  We demonstrate how the public can be systematically mislead even in those circumstances where all experiments are published and scientists exhibit no bias in favor of industry.}.

An important take-away from these arguments is that what may seem like weak or subtle interventions are often most effective for the would-be propagandist.  In particular, outright scientific fraud---intentional publication of incorrect, fabricated, or purposely misleading results---is not only unnecessary to influence public opinion on topics of scientific inquiry, it is also riskier and often less effective than other forms of manipulation.  Biased production, which does not involve fabricating results, is a successful strategy for misleading the public. And in many cases, biased production is itself less effective than selective sharing---even though it involves more direct and explicit interventions on the production of science.  These observations support claims by \citet{holman2017experimentation} that more subtle industry influences---like their ``industrial selection'', which we will discuss at length in the conclusion of this paper---can be very effective.  In fact, as we will also argue in what follows, even interventions with no malicious intent, such as those of journalists who aim to be ``fair'' in their reporting on controversial science, can unwittingly mimic the effects of an active propagandist.

Before proceeding, we wish to emphasize what is at stake in this discussion. Recognizing how the Tobacco Strategy works is essential for understanding the relationship between science, public policy, and democracy.  There is now a large literature in philosophy of science on what it means for science to be suitably ``democratic,'' in the sense of reflecting the needs and values of the members of a society affected by scientific results.\footnote{See, for instance, \citet{LonginoSSK,LonginoFK}, \citet{KitcherSTD,KitcherSDS}, \citet{DouglasSPVFI}, \citet{ElliottLPGFY}, and references therein.}  This literature focuses on how democratic ideals should shape scientific research.  But the other direction, of how to best understand the role that scientific knowledge plays in democratic institutions, is equally important---and requires us to reckon with the fact that there are powerful political and economic interests that wish to manipulate and interfere with the distribution of scientific knowledge.  Understanding the mechanisms by which public understanding of science can fail is crucial to developing institutions and interventions which might be effective in defanging the Tobacco Strategy, and, in the case of `fair' reporting, improving journalistic practice to protect public belief.

The remainder of the paper will be structured as follows.  In the next section we will introduce the model we consider in the paper.  Sections \ref{sec:selectiveSharing} and \ref{sec:biasedProduction} will present results concerning selective sharing and biased production, respectively.  Section \ref{sec:journalism} will describe results from a slightly modified model in which we consider ways in which journalistic practice with no malicious intent can simulate the actions of a propagandist.  We will conclude with some reflections on the results, including a discussion of possible interventions.

\section{Epistemic Network Models}

\citet{venkatesh1998learning} introduce a model where individuals in a community try to choose which of two theories, or practices, is the better one.   In order to decide, they make use evidence from two sources.  First, each individual tests the theory they favor, and second, they share their results with those in their social network.  Over time, they update their beliefs about the theories based on this evidence, until, in many cases, the whole group comes to believe that one or the other is better.  \citet{zollman2007communication,zollman2010epistemic} shows how simulations of this model can be applied to understand scientific discovery and consensus.  Under his interpretation, each individual in the model is a scientist testing a hypothesis.  For instance, suppose a new treatment for cancer is introduced.  How do researchers go about deciding whether to adopt the new treatment or stick with the old one?

The model starts with a network, defined by a set of scientists and a set of fixed links representing pathways of communication between them.  Inquiry is modeled using what is called a two-armed bandit problem.  Each scientist is aware that there are two actions they can perform.  Action A is well understood, and is known to be successful 50\% of the time.  Action B, on the other hand, is less well understood, and the agents are uncertain whether it is more or less successful than A.  In particular, the instantiation of the model we use here assumes that B is a bit better than A, so that its chance of success is $p_B =.5 + \epsilon$.\footnote{The payoff of success for both actions is the same, and failure yields no payoff, so action B has a strictly higher expected payoff.}  Scientists, however, do not know whether it is a bit better, or a bit worse (success rate $.5 - \epsilon$). Their aim is then to discover which of these possibilities is the better one, and so adopt the superior action.  In what follows, we will refer to ``theory A'' as the view that action A is preferable, and ``theory B'' as the view that action B is preferable.

At the beginning of a simulation, scientists are given random credences about which theory is better, represented by numbers between 0 and 1.  (For example, a scientist with credence .7 believes there is a 70\% chance that theory B is right.)  Each round, scientists apply their preferred theory to the world.  Agents with a credence $< .5$ believe that theory A is probably correct and employ it.  Agents with credence $> .5$, on the other hand, use theory B.  This might correspond to doctors who approve of a new treatment deciding to try it on patients, while others stick with a more conservative approach.

In each round, scientists observe how their choices worked, and change their beliefs in light of new evidence.  In particular, it is assumed that they test their theory some number of times, $n$. They then update their credence using Bayes' rule based on what they, and all their network neighbors, have observed.  Simulations stop when either all agents believe in theory A---and so do not learn anything new about theory B---or when all agents have reached some certainty threshold (say .99 credence) that B is correct, and so are exceedingly unlikely to end up changing their minds.  Both outcomes occur in simulations.  Sometimes initial beliefs and/or strings of data supporting theory A are enough to create a mistaken consensus around the worse theory, and sometimes the community gets enough good data to settle on the better one.

As noted in the introduction, we are particularly interested in the role of industry, and propagandists in general, in shaping policy based on science, and in influencing popular belief.  For this reason, our models follow the framework developed by \citet{venkatesh1998learning} and \citet{zollman2007communication}, but with a few, key alterations.  First, we introduce a new set of agents, which we will call policy makers.  These agents have beliefs, and update these in light of evidence, but do not gather evidence themselves.  Each one listens to some, or all, of the scientists in the network, but does not influence them or anyone else. We assume that at the beginning of each simulation, policy makers are skeptical about the new, uncertain theory, in the sense that their initial credences are sampled from a uniform distribution supported on the interval $(0,.5)$.

We also introduce one further agent that we call the propagandist.  This agent observes all the scientists, and communicates with all the policy makers.  Their goal is to convince the policy makers to prefer the worse theory A, by sharing evidence that suggests action A is preferable to action B.  We consider two treatments, corresponding to different strategies by which the propagandist may do this. The idea is to represent an individual, or set of individuals, who are motivated to use all their resources to publicize data that is ultimately misinformation.  The actions of the propagandist capture various aspects of the Tobacco Strategy: by sharing a biased sample of evidence with policy makers (and the public) they seek to divide public opinion from scientific consensus.\footnote{\label{howpossibly} It is common in discussions of idealized models of epistemic communities---and philosophical discussion of modeling in the social sciences generally---to suppose that such models provide, at best, ``how-possibly'' information about a target phenomenon. (See \citet{frey2018epistemic} for a discussion of this view.) We wish to explicitly reject this attitude, at least with regard to the inferences we draw from the present model.  On our view, there is ample and convincing historical evidence that the tobacco industry has intervened to influence public beliefs using strategies very similar to the ones discussed here.  The present model supplements this historical evidence by allowing us to ask counterfactual questions concerning how precisely this kind of intervention works.  For this reason, we adopt the position that our model provides ``how-potentially'' information concerning the causal role of this sort of intervention.  What we mean by this is that the model is realistic enough to explore processes that have the potential to really occur. In this sense, we take this model to lie somewhere on a spectrum between ``how-possibly'' and ``how-actually'' (i.e., fully realistic) models. Exploring how-potentially models can direct further inquiry into whether at these processes are really at play, or (as in this case) supplement existing empirical work on these processes (\citet{rosenstock2017epistemic}).   See also \ref{causal}.  We are grateful to an anonymous referee for pressing us to clarify our stance on these issues, and for emphasizing that these models lie on a spectrum (a point with which we agree).}

\section{Selective Sharing}
\label{sec:selectiveSharing}

In the first treatment, which we call \emph{selective sharing}, the propagandist adopts the following strategy: in each round of the simulation, they search the network of scientists for results that tend to favor theory A, and then they share only those results with the policy makers.  The policy makers then update their credence in light of both the evidence they receive from the scientists to whom they are connected and the evidence received from the propagandist, which includes a subset of the results produced by the scientists.\footnote{As we discuss below, this model involves some double-counting of results favorable to theory A, as a given policy maker may receive the same evidence from both a scientist and from the propagandist.}  In other words, each policy maker receives, and updates on, a biased sample of evidence including a disproportionate number of results favoring theory A.  However, all of the results the policy makers receive are ``real'' in the sense that they are produced by members of the scientific community, without bias or influence by the propagandist or anyone else.  In addition, the propagandist does nothing to influence the scientific community; they only seek to shape policy maker beliefs.\footnote{In this way, the model we consider differs significantly from that of \citet{holman2015problem}, who consider an industry scientist who influences other scientific beliefs.  We will return to this relationship in the next section, as the treatment discussed there is somewhat closer to that discussed by Holman and Bruner.}

\begin{figure}[h]
\begin{center}
\includegraphics[width=.5\columnwidth]{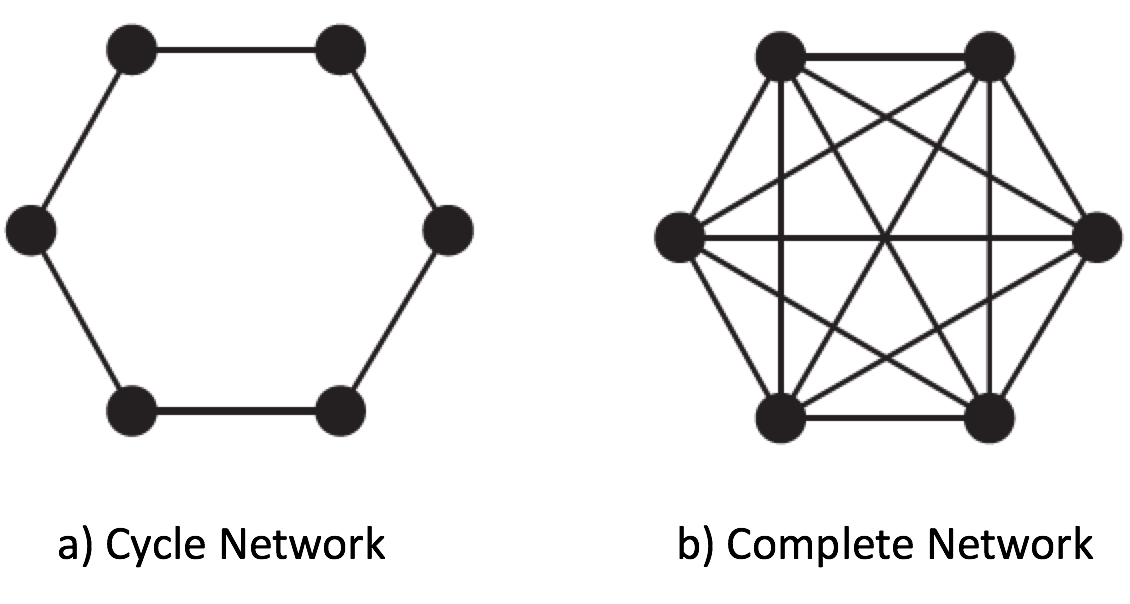}\end{center}
\caption{Cycle and complete network structures.}
\label{fig:Networks}
\end{figure}

We analyze this model by varying parameter values including the number of scientists in the network, $K$, the number of draws each round, $n$, the probability that action B is successful, $p_B = .5+\epsilon$, and the number of scientists each policy maker listens to, $k$.\footnote{We consider subsets of the following parameter values: population size $K = 4, 6, 10, 20, 50$, number of draws $n=1, 2, 5, 10, 20, 50, 100, 500, 1000$, success of better theory $p_B = .51, .55, .575, .6, .7$, and number of scientists listened to by each policy maker ranges from 1 to all of them.  Our choices were influenced, to some degree, by tractability considerations; simulations with low $p_B$ and many scientists, in particular, took longer to run.}  (Each policy maker could connect to one scientist, or to two, or to three, etc.)  We consider models where the network of scientists is either a cycle, where each scientist is connected to two peers in a ring structure, or a complete network, where every scientist communicates with every other.  Figure \ref{fig:Networks} shows these two structures.  They differ in that it will tend to take longer for the cycle to converge to a consensus, especially for large communities, because fewer scientists share data with each other. This network structure affects our results, as we will make clear, mainly by influencing the amount of time it takes the communities of scientists to reach consensus.  All reported results were averaged over 1k runs of simulation that terminated when either correct or incorrect consensus was reached.\footnote{We ignore outcomes where scientists do not converge to the correct belief.  (In these cases, policy makers also tend to accept the worse theory.)  This means that our results are averaged over fewer than 1k rounds.  In the worst case scenario, though, this only meant ignoring~10\% of data points.}

We find that the presence of a single propagandist who communicates only actual findings of (ideal, unbiased) epistemically motivated scientists can have a startling influence on the beliefs of policy makers.  No one in this scenario is doing bad science.  No one is falsifying data, or even publishing it selectively.  The propagandist communicates only real scientific findings.  Nonetheless, under many scenarios we find that while the community of scientists converges on true beliefs about the world, the policy makers reach near certainty in the false belief.  This point will become clearer as we share results in this section and the next---for a wide range of parameter values, public belief is highly accurate when no propagandist is at play, but as inaccurate as it could possibly be under the influence of the propagandist.

\begin{figure}[h]
\begin{center}
\includegraphics[width=.8\columnwidth]{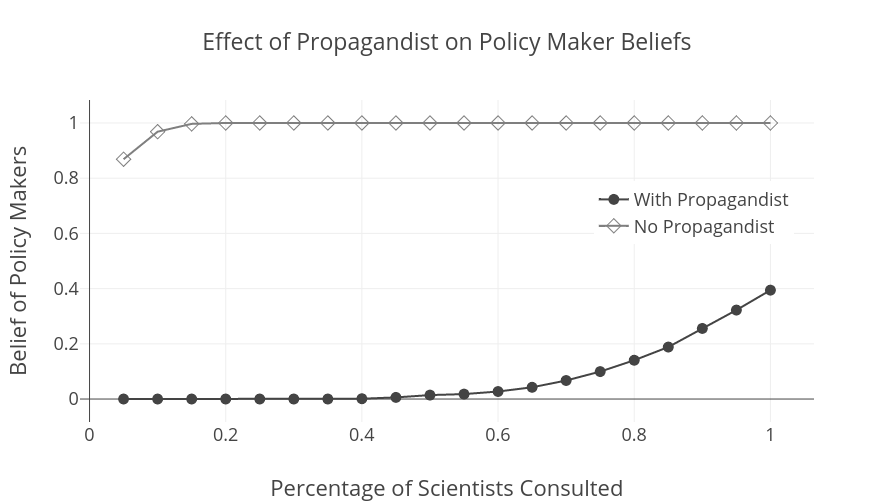}\end{center}
\caption{Policy maker beliefs as influenced by their connections to scientists.}
\label{fig:numbersci}
\end{figure}

We find that the likelihood of this sort of disagreement between science and public belief depends on features of the community and the problem studied.  First, unsurprisingly, when policy makers listen to more scientists they are harder to deceive.  Figure \ref{fig:numbersci} shows the average credence that policy makers have in the correct theory at the point where scientists reach a correct consensus, which we understand as a state in which all members of the scientific community have credence greater than .99.  The x-axis tracks the proportion of scientists in the entire community that each policy maker consults.  As we can see, without a propagandist, public beliefs lag behind scientific beliefs slightly, though the more scientists are consulted, the better the public does.  With a propagandist, the public is convinced that the false theory is preferable, though if they are connected to more scientists, this effect is ameliorated.\footnote{These results are a cycle network with $K = 20$, $n = 10$, and $p_B = .55$.  This trend occurred across all parameter values, as did the other trends reported unless otherwise specified.}  In related work, \citet{holman2015problem} model the effects of industrial propagandists who share biased data with \emph{scientists} and observe that it is best for scientists to be highly connected to each other in such cases.\footnote{This shows how the suggestion from \citet{zollman2007communication,zollman2010epistemic} that less connected scientific communities are more successful at reaching the truth does not hold when industry attempts to sway consensus.}

\begin{figure}[h]
\begin{center}
\includegraphics[width=.8\columnwidth]{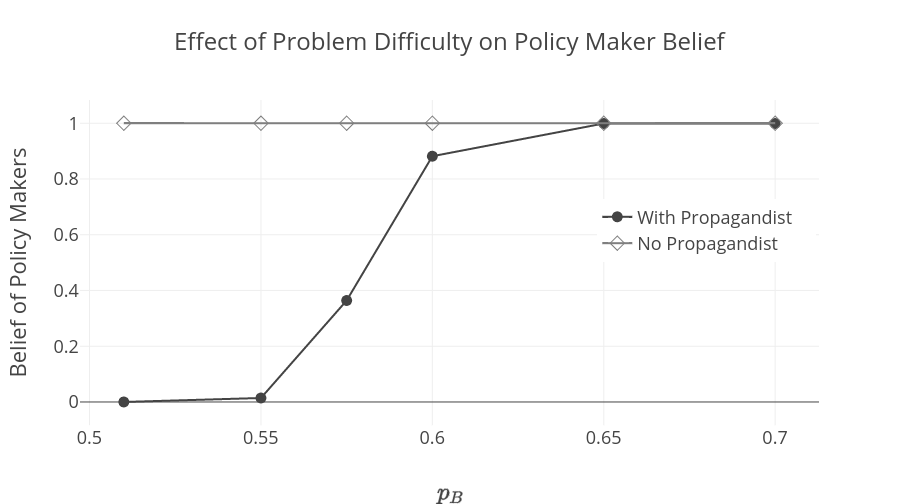}\end{center}
\caption{Policy maker beliefs as influenced by the difficulty of the scientific problem.}
\label{fig:Difficulty}
\end{figure}

It is also harder to deceive policy makers when the size of the effect---i.e., the difference in expected payoff between action A and B---is larger.  As $p_B$ (the probability that the better action pays off) increases, so that the difference between the better and worse theories is more stark, the policy makers are more likely to learn the truth.  This is because there are fewer studies that happen to support the worse theory, and thus less material for the propagandist to publicize.  Figure \ref{fig:Difficulty} shows how as the difference between the success rates of the two theories increases, policy makers tend to end up holding the true belief regardless of propaganda.\footnote{Results are for $K = 20$ scientists in a cycle, each policy maker listening to 50\% of scientists ($k = 10$), and $n=10$.  Again the trend is stable across parameter values.} This observation is especially relevant to the biomedical sciences, because human bodies are highly variable, which means that data is often highly equivocal.  We should expect studies of human subjects to provide plenty of material for potential propagandists looking to obscure true evidential trends.

\begin{figure}[h]
\begin{center}
\includegraphics[width=.8\columnwidth]{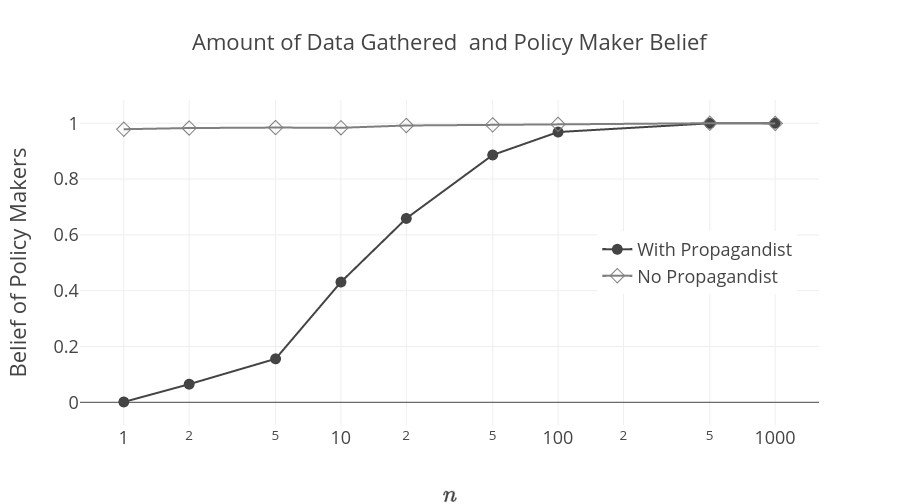}\end{center}
\caption{Policy maker beliefs as influenced by the amount of data gathered by scientists.}
\label{fig:amtdata}
\end{figure}

The number of draws each round, $n$, also has a strong effect on the ability of the propagandist to mislead.  This parameter may be thought of as the sample size for each experiment performed by the scientists.  For fixed $p_B$ (i.e., fixed effect size), it is roughly analogous to the statistical power of each scientist's experiments: the higher it is, the less likely that a study will erroneously support theory A.\footnote{More precisely, if we take the null hypothesis to be that that $P_B < P_A$, and the alternative hypothesis to be that $P_B > P_A$, then for virtually any hypothesis test one considers, if we fix the effect size, $\epsilon = |P_B - .5|$,  increasing $n$ will increase statistical power in the sense that it will increase the probability that the false null hypothesis is rejected in favor of the true alternative hypothesis.  Of course, our agents are Bayesians, and so frequentist notions such as power are not strictly relevant---the agents update their beliefs in light of any and all evidence that they receive.  Still, thinking of $n$ as roughly tracking power is useful, here, because it gives an intuitive sense of why small $n$ studies are generally better for the propagandist.}  Thus scientists who run the equivalent of well-powered studies and produce fewer spurious results also produce less for the propagandist to work with.  Figure \ref{fig:amtdata} shows the final average beliefs of policy makers with and without a propagandist as $n$ increases.\footnote{This figure is for $K = 10$ scientists in a complete network, each policy maker listening to 90\% of scientists, and $p_B = .55$.}   Note that the x-axis is on a log scale to make the trend more visible.  For high enough $n$, the propagandist can do almost nothing since they very rarely have the spurious results necessary to mislead the public.  In these cases, policy makers figure out the truth despite the propagandist's efforts.\footnote{In an interesting connection, \citet{holman2015problem} find that when a propagandist shares biased data with scientists, and when these scientists are able to ignore data that is statistically unlikely given their credences, scientists do better when $n$ is large.  This is for a different reason than we describe.  When $n$ is large in their models, it is easier to detect a propagandist and ignore them.  If $n$ is small, more results are statistically likely given some agent's credence.}

One might think that increasing the number of scientists in a network, $K$, holding all else fixed, would decrease the effectiveness of the propagandist.  More scientists means more data, and more epistemically motivated minds searching for truth.  But this does not occur in general in these models.  Assume that policy makers listen to some fixed number of scientists, $k$.  This is not an unreasonable assumption given limits on the amount of time and energy policy makers or the public devote to science.  In such cases, as the total number of scientists increases, there is no benefit to policy makers, since they continue to listen to $k$ scientists.  On the other hand, the increased number of scientists increases the total number of spurious results that the propagandist can then share with policy makers, often worsening policy maker belief.\footnote{This is not always the case since as we outline in the Appendix the size of the community also influences the speed of convergence to scientific consensus, which can influence policy maker beliefs.}  In the next section, we will engage in some further discussion of the effects that the size of the scientific community has on policy maker beliefs.

The final consideration that influences the success of the propagandist has to do with how much time it takes for the scientific community to converge to consensus.  The convergence time depends on various factors.  For instance, the size of a network generally influences how long it takes for all scientists to reach agreement.  (In a sparsely connected network, more scientists means a longer time until each one is convinced.)  In addition, as mentioned above, for fixed population size the cycle and complete network arrangements differ in that cycle networks take longer to converge.  This is because scientists listen to fewer of their colleagues, meaning that disagreements will tend to persist as actors in different parts of the network gather different sorts of data.

Previous results from \citet{zollman2007communication,zollman2010epistemic} have shown that in many cases factors that slow convergence can be beneficial for an epistemic community.  These factors lead to transient diversity of opinion, which is important for preventing scientists from settling too quickly on a theory that might be wrong.  Heavily connected communities, on the other hand, are easily swayed by strings of misleading evidence that are widely shared.\footnote{\citet{rosenstock2017epistemic} show that this will only occur for the region of parameter space in which the problem of selecting the better theory is hard (due to there being fewer scientists, less difference between theories, or lower $n$).}  But our results show that other considerations can complicate this issue.  When industrial actors have a stake in a community's findings, taking a long time to settle on consensus can provide more opportunity for the propagandist to find and promote results that went the wrong way under conditions where spurious results are likely.

To try and better understand the relationship between time to convergence of the scientific network and the effectiveness of the propagandist, we derive semi-analytic results predicting expected outcomes in this model.  In general, we can think of each policy maker's belief as influenced by two separate factors.  One comes from the community of scientists, which, on average, will tend to pull credences in theory B up.  The other comes from the propagandist and will tend to pull credence in theory B down.  In other words, a policy maker's belief in round $t+1$ of simulation on average can be represented by:
\[B_{(t+1)} = Y_P \times Y_S \times B_t,\]
where $B_t$ is their prior belief, and $Y_P$ and $Y_S$ represent the average effects of the propagandist and scientists.  In the appendix, we derive expressions for these quantities and show that it will always be the case that $Y_P <1$ and $Y_S >1$.

We can thus think of the dynamics in this model as a sort of tug of war between science and industry over public belief.  In general, for any set of parameter values in our model, and some state of scientist/policy maker beliefs, one side will be winning because either $Y_P \times Y_S <1$ or $Y_P \times Y_S >1$.\footnote{The situation in which the two influences exactly cancel is a knife's edge case.} This means that when the propagandist is winning on average, they do better during long periods of scientific debate. In such cases, policy makers do better when scientists quickly reach consensus.  Conversely, when the $Y_P\times Y_S > 1$ more evidence will tend to to lead the policy makers towards the true belief---with the propagandist's main effect being to slow this progress.\footnote{One interesting observation is that the size of the scientific network, besides influencing time to convergence, also influences the rate at which policy makers change their beliefs.  Adding more scientists increases the total amount of data both coming directly from scientists and coming from the propagandist.  When the scientists are winning the tug of war, this increase benefits the policy makers.  When the propagandist is winning, this simply increases the rate at which policy makers decrease their credence in $P_B$.}

\begin{figure}[h]
\begin{center}
\includegraphics[width=.8\columnwidth]{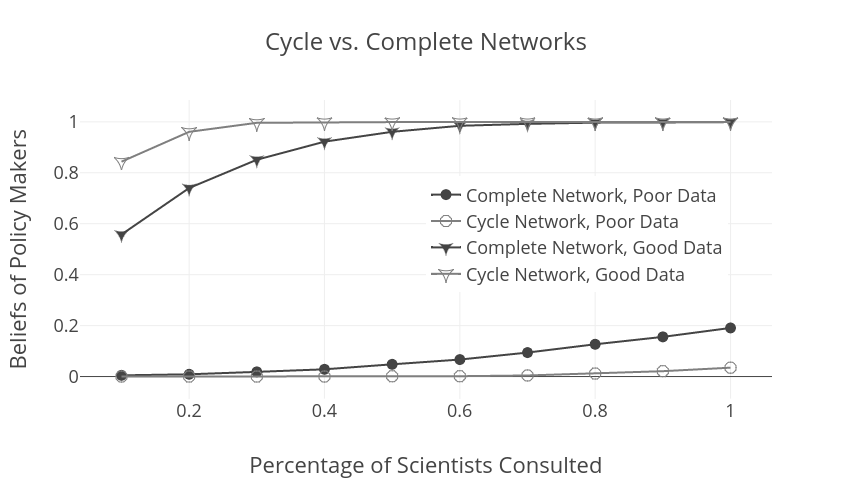}\end{center}
\caption{Influence of scientific network structure on policy maker beliefs.}
\label{fig:cyclecomplete}
\end{figure}

For an example of the effects of time to convergence on belief, consider figure \ref{fig:cyclecomplete}, which shows the beliefs of policy makers consulting cycle and complete networks.  We show results in a scenario that favors true belief---$p_B = .7$ and $n = 20$---and also one that disfavors it---$p_B = .55$ and $n = 5$.  In the first case, the slower cycle network gives policy makers time to learn the correct theory from scientists, while relatively unencumbered by the propagandist.  In the second case, the faster complete network gives the propagandist less time to share spurious data, and so is better for policy makers.

To be clear, if we look again at the simulation results we have presented, we can observe that, since policy makers, on average, start simulation at credence .25, any outcome where their final beliefs are $<.25$ is one where the propagandist is pulling harder than the scientists, and vice versa.  If we imagined that the process kept going---scientists kept performing action B, and the propagandist kept publicizing spurious results---we would expect policy makers to become completely convinced in either A or B eventually, with the relative values of $Y_S$ and $Y_P$ determining which one.  But the community can end up at a better or worse state depending how quickly this happens.  The most important take-away here is that there are reasons to think that a long argumentative process between scientists might actually be a bad thing when public belief is on the line.  Thus we find a curious trade-off with the Zollman effect: at least in some cases, maximizing convergence time is truth conducive for the scientists, but not for the public.

In the next section, we will move on to discuss the second part of the Tobacco Strategy---where industry funds their own scientists and disseminates the resulting research.  Before doing so, we mention a few potential concerns about the models just discussed, and robustness checks meant to alleviate those concerns.  First, as noted above, data in our model can be ``double counted'' in the sense that a policy maker might update their belief on data from a scientist and then update on the same evidence as it comes from the propagandist.  We take this to be a feature, rather than a bug, because it captures how the prevalence and repetition of evidence can increase its impact.  Nonetheless, we also ran the same simulations where data could not be double counted this way.  Results were qualitatively similar, though the strength of the propagandist was diminished, particularly for policy makers connected to many scientists.  We also varied the pattern with which policy makers connected to scientists to ensure that this did not have a significant effect on the results.

\section{Biased Production}
\label{sec:biasedProduction}

As mentioned in the introduction, another part of the Tobacco Strategy involved producing industry-funded data that could be used to counter the research coming from the scientific community.\footnote{There is a close connection here to the work of \citet{holman2015problem}, who consider biased agents in a scientific community.  Like our ``biased production'' propagandist, their ``intransigently biased agent'' (IBA) produces and shares evidence that tends to favor the wrong theory.  However, in their case the IBA draws from a biased arm relative to the other scientists in the community, whereas, as we describe, our propagandist performs the same actions as the scientists but selectively publishes results.  In addition, they focus on the effect an IBA might have on beliefs within a scientific community, showing how industry can shape scientific consensus itself, rather than public belief in particular.}  Our second treatment concerns this aspect of the strategy.  To model this, we now assume that the propagandist has their own resources that they can allocate towards gathering data about the uncertain theory B.  After gathering this data, the propagandist then shares only those studies that spuriously indicated that theory A was the better one.  Notice that we do not allow the propagandist to do `bad science' in the sense of falsifying data, or to use biased methodologies.  Their data gathering process is, in fact, identical to scientists attempting to find the truth.  The only intervention we allow on the scientific process is one that is currently widely practiced (and not always considered particularly unethical): selectively choosing not to publish some results.

We varied several new parameter values in this treatment.  First, we varied the total amount of resources available to the propagandist.  (Think of this as the total number of pulls from a bandit arm the propagandist can perform.)  We also varied how the propagandist chose to divide up these resources into studies.  Did they run many low powered studies, or just a few high powered ones?\footnote{We ran simulations for a subset of the following parameter values: population size $= 4, 6, 10, 20$, number of draws $n=1, 2, 10, 25, 50, 100$, success of better theory $p_B = .55, .6, .7$., and number of scientists consulted, $k$, ranged from 1 to all of them.  Propagandist total resources varied from equal to that of one scientist, to equal to those of the whole scientific community.  Propagandist $n$ ranged between $1$ and their total possible pulls.}

In this scenario, holding other parameters fixed, more scientists are unequivocally beneficial to policy makers' beliefs.  Scientists no longer unintentionally provide grist for the propagandist, and, instead, produce and publicize data that is more likely to point in the right direction.  It is also consistently better, as in the last section, for policy makers to consult more scientists.  The more data they receive from epistemically motivated actors, the less the impact of the propagandists' spurious studies on their beliefs.  And, again, the better the successful theory, the lower the chance that the propagandist generates enough spurious results to influence policy makers.

\begin{figure}[h]
\begin{center}
\includegraphics[width=.8\columnwidth]{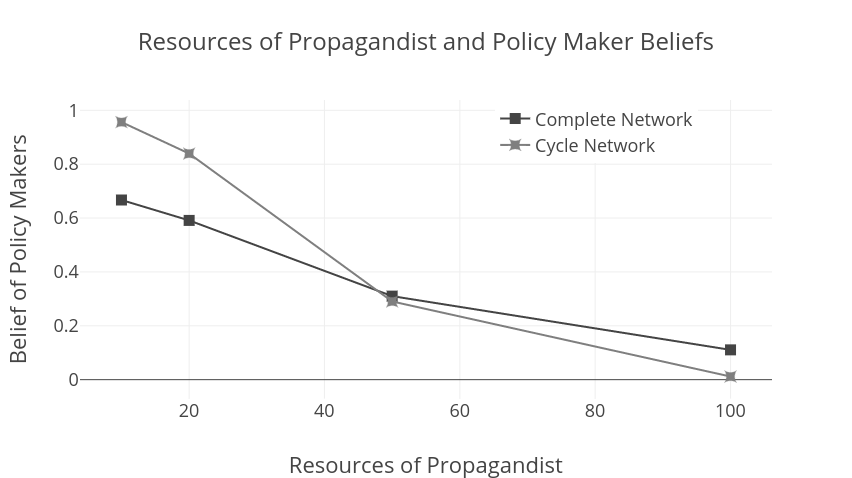}\end{center}
\caption{Propagandists with more resources are better able to mislead policy makers.}
\label{fig:resourceprop}
\end{figure}

Unsurprisingly, more resources are better for the propagandist.  If they have the wherewithal to run more studies, they have a greater chance of generating results supporting theory A.  Suppose each scientist has the resources to collect 10 data points per trial.  Assume that the propagandist runs trials of the same size, but can run a different number of these depending on their level of resources.  As figure \ref{fig:resourceprop} shows, as this number increases, the likelihood that the policy makers believe the correct theory goes down.\footnote{These results are for 10 scientists, $n = 10$ for scientists, $p_B = .55$, and policy makers listening to half of scientists.}

Suppose that we hold fixed the resources a propagandist has at their disposal.  The last question we ask is: given this, what is the best way for the propagandist to allocate these resources to different studies?  The answer is that it is always better for them to break their data up into the smallest possible chunks.  Suppose the propagandist has resources to collect 1000 data points.  In the most extreme case, we could imagine a scenario where the propagandist runs 1000 studies with one draw each.  They could then report only those where action B failed (on average this would be $(.5-\epsilon) * 1000$ data points), making it look to policy makers as if action B was never successful in these experiments.  At the other extreme, the propagandist would run a single experiment with 1000 data points.  The chances that this experiment would support the better theory B are quite high, leaving the propagandist with nothing to disseminate.  In general, the propagandist should choose the smallest $n$ they can get away with.

\begin{figure}[h]
\begin{center}
\includegraphics[width=.8\columnwidth]{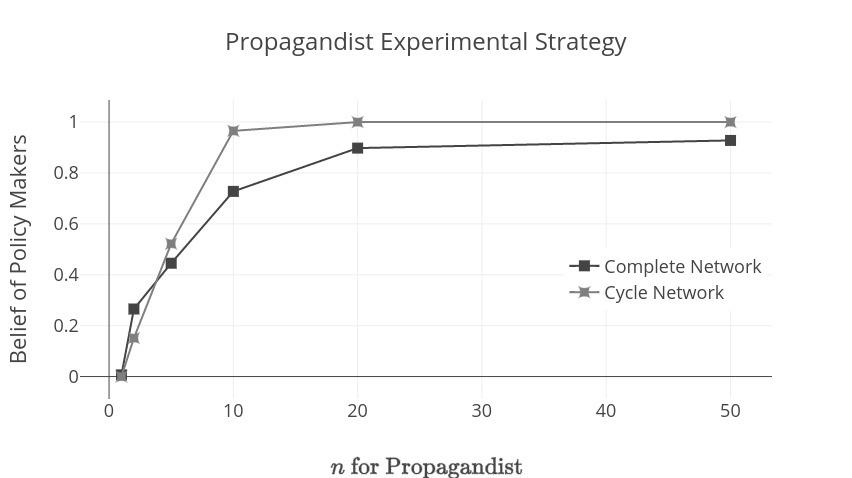}\end{center}
\caption{Propagandists who use their resources to perform small studies are better at misleading policy makers.}
\label{fig:propstrat}
\end{figure}

Figure \ref{fig:propstrat} shows this effect.  For propagandists with resources to gather 50 data points, the smaller the studies they run, the worse the eventual beliefs of the policy makers.  If propagandists run a few, good studies, they are unable to mislead anyone.\footnote{These results are for 10 scientists, $n = 10$ (for scientists), policy makers listening to all scientists, and $p_B = .55$.}

\begin{figure}[h]
\begin{center}
\includegraphics[width=.8\columnwidth]{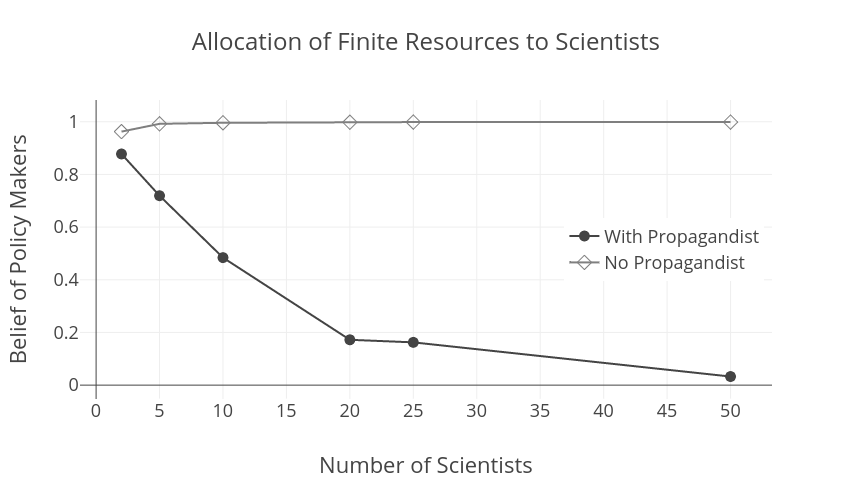}\end{center}
\caption{Policy makers do better when fixed resources are allocated to a few, good studies rather than many with small samples.}
\label{fig:resourcealloc}
\end{figure}

This observation actually has an analog in the previous set of results on propagandists who selectively share others' data rather than gathering their own.  In that model, given a set of available resources for a community, it is better to use these to run fewer, better studies than to run many less well-powered ones.  We can observe this by comparing how policy makers do given communities that have a finite amount of resources distributed differently.  Figure \ref{fig:resourcealloc} demonstrates this.  The scientific network, in this figure, has the resources to gather a total of 100 data points each round.  This is divided up by the number of scientists involved in exploring the phenomenon so that, for example, 10 scientists gather 10 data points each or two scientists gather 50 data points each.  Without a propagandist policy makers develop successful beliefs regardless of how resources are allocated.  As the individual studies get smaller, though, the propagandist is increasingly able to take advantage of the stochasticity involved in gathering data to forward their agenda.  Small studies mean confused policy makers.

\section{Journalists as Unwitting Propagandists}\label{sec:journalism}

In the two treatments we have discussed so far, the propagandist has adopted strategies intended to influence the policy makers' beliefs.  In both cases, we found that the propagandist can succeed, at least under some circumstances.  The basic mechanism at work in both strategies is that the propagandist endeavors to bias the total body of evidence that the policy makers see.  They do not do this through fraud, or even by sampling from a different distribution than the other scientists (as in \citet{holman2015problem}).  Instead, they do it by choosing results so that the total evidence available, even though it was all drawn from a fixed distribution, does not reflect the statistics of that distribution.   What these results show is that in many cases, particularly when evidence is relatively difficult to get or equivocal, it is essential that policy makers (and the public) have access to a complete, unbiased sampling of studies.  This suggests that curation, or even partial reporting, of evidence can lead to epistemically poor results.

But propagandists are not the only agents who regularly curate scientific results.  Non-malicious actors can also play this role.  For instance, journalists often select scientific studies to feature in articles.  Few journalists are experts in the relevant science, and in any case, the goal is rarely to give a uniform and even-handed characterization of all available evidence.  To the contrary, journalistic practices often encourage, or even require, journalists to share either only the most striking, surprising, or novel studies; or else, when they choose to share a study on a controversial topic, to endeavor to be ``fair'' by sharing research that supports both sides.  Indeed, the U.S. Federal Communications Commission (FCC) had an official policy, from 1949 to 1987, that \emph{required} broadcast news to present controversial topics of public interest in a ``balanced'' manner.  In practice, this meant giving roughly equal time to multiple perspectives.  The policy was known as the ``Fairness Doctrine''. \citet{Oreskes+Conway} argue that this Doctrine became a tool of the tobacco industry's propagandists, who used it to insist that pro-industry science get publicized.

To study the effects of the Fairness Doctrine, we consider the following modification of the models we have considered thus far.\footnote{We also considered a version of the model in which the journalist only published results that were sufficiently ``surprising'' in the sense that the journalist judged them to have had a low probability of occurring.  For some parameter values, this sort of journalistic practice led policy makers to accept the wrong theory.  But we deemed the results to be ambiguous, because they depended sensitively on the value of $n$ chosen.}  The network structures remain the same, but in place of the propagandist, we have a ``journalist'', who like the propagandist can see all of the scientists' results and can communicate with all of the policy makers.  In this case, we suppose that the policy makers receive all of their information from the journalist, and consider different ways in which the journalist might share that data.

Suppose, for instance, that the journalist---abiding by the Fairness Doctrine---shares two results from the network of scientists each round, one of which supports theory B, and one of which supports theory A.  More precisely: we suppose that each round, the journalist chooses randomly among all of the results that support theory A and all that support theory B.  In rounds where no results support one of the two theories, the journalist shares the result they shared the previous round (capturing the idea that sometimes journalists need to search past data to find something to adequately represent the ``other side''), unless there has never been a result favoring a given theory, in which case the journalist shares only one result.  What we find is that this addition artificially slows the policy-makers' acceptance of theory B.

To show this, it is most informative to compare the policy makers' beliefs under the action of this journalist (Fair) with what one would find if (a) the journalist shared two randomly selected pieces of data each round (Random) and (b) the journalist simply shared \emph{all} available evidence each round (All).  We find that (on average across runs) the Fairness Doctrine never improves the epistemic performance of the policy makers, and for many parameter values, policy makers receiving evidence from a journalist abiding by the Fairness Doctrine end up with beliefs substantially less accurate than under conditions (a) or (b).  In all cases we consider, the policy makers' beliefs still tend towards the truth (whenever the scientists' beliefs do), but they do so much more slowly.\footnote{When policy makers see one result supporting B and one supporting A each round, on average the result supporting B is still stronger since this is in fact the better arm.  So they learn the truth eventually.}  In this sense, then, the ``fair'' journalist mimics the propagandist under the conditions where they are losing the tug-of-war for policy makers' beliefs, but nonetheless slow the policy makers' progress.

\begin{figure}[h]
\begin{center}
\includegraphics[width=.8\columnwidth]{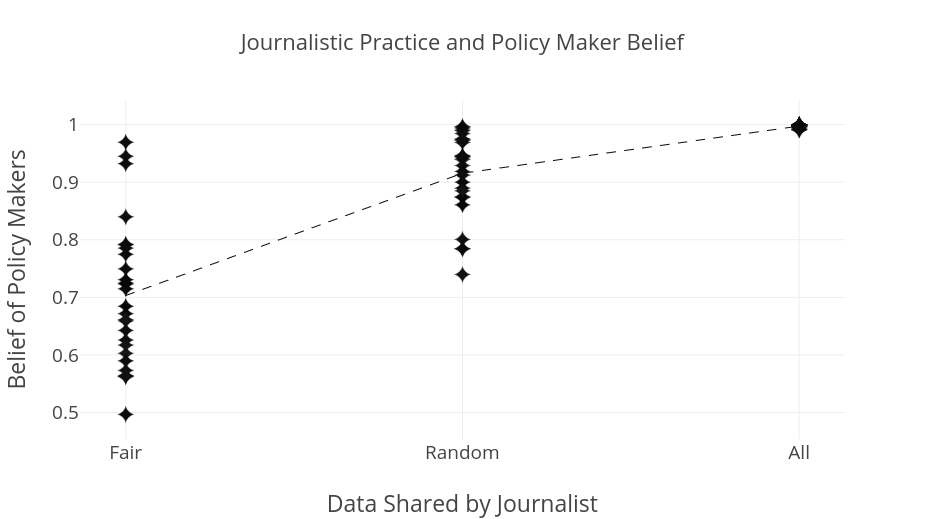}\end{center}
\caption{`Fair' reporting slows policy makers from reaching accurate beliefs.}
\label{fig:journalist}
\end{figure}

Figure \ref{fig:journalist} shows average policy maker beliefs when scientists reach consensus on the true belief for complete networks under different parameter values.\footnote{We chose a representative set of data points including all combinations of $K = 6, 10, 20$, $p_B = .55, .6, .7$, and $n = 5, 10, 20$.  As in the previous models, initial policy maker beliefs were skeptical.  Complete networks demonstrated the effect more strongly because cycle networks are slow to converge and thus there is more time for policy makers to learn true beliefs even under `fair' reporting.  But even with cycle networks the effects of fair reporting were always neutral or negative.}  Data points are for simulations where journalists share Fair, Random, or All data as described above.  The trend line shows the average beliefs across parameters under each condition.  As is evident, when journalists share all the data, policy maker beliefs are very accurate.  With less data available---two random sets of data---their convergence to true beliefs is slowed.  Under fair reporting, policy makers converge to true beliefs much more slowly, since they see a result that supports theory A every time they see a result supporting B.  The strength of this effect is dependent on parameter values, but, as mentioned, fair reporting only worsens, and never improves, policy maker beliefs in comparison to reporting randomly or reporting all data.

Journalists are not the only actors who perform data curation. University faculty, when teaching courses, may endeavor to give a panoptic and even-handed view of a field, but they will not be able to share all available relevant data with a class.  Literature reviews and survey articles also curate data, commenting on its quality and selecting what seems most relevant or significant.  The mechanisms by which researchers and university instructors select material to present are arguably motivated by epistemic considerations---they seek to give a compact overview that supports the views they believe are best supported, given the total evidence that they have seen---but it is hard to see how they can do this without using their own beliefs---which may be influenced by various factors---to guide the choices.  And so, it is not clear that even this sort of curation is certain to accurately track the statistics of the total available evidence.  In this way, even independent researchers, in their teaching and writing, can unwittingly act as propagandists.  In addition, it has been widely observed that scientists regularly fail to publish negative or null results.  This `file drawer' effect biases the pool of evidence available to the entire community, and has already been shown to potentially have negative effects on scientific beliefs (\citet{romero2016can}).\footnote{See also \citet{ioannidis2005most}.}

\section{Conclusion}

We have now considered two aspects of the Tobacco Strategy as described by Oreskes and Conway.  We take the results we have presented to provide strong evidence that the Tobacco Strategy can work, in the sense that biasing the total evidence presented to a group of agents can lead those agents to converge to the incorrect belief.  Of course, modeling alone cannot establish that this \emph{did} happen in the case of tobacco industry propaganda efforts.  But when you combine the modeling results we have presented with the careful historical analysis and arguments offered by \citet{Oreskes+Conway}, there are strong reasons to believe that the Tobacco Strategy was an important causal factor in the long delay between when a link between cigarettes and cancer was established by the medical research community, and large-scale changes in regulation and public attitudes.  Moreover, we take it that if these strategies have succeed in the past, they are very likely to succeed now and in the future.\footnote{Returning one more to the themes of notes \ref{causal} and \ref{howpossibly}: we emphasize that, at very least, these models provide ``how-possibly'' information: the processes we decribe with these models can possibly occur.  Furthermore, as we argue here, the theoretical and historical knowledge developed by Oreskes and Conway increases our confidence that these processes have been important ones in the real world.  As noted, we cannot be sure that this is the case.  But we do not take this to be a special problem for models of this sort.  With almost any sort of scientific exploration, including experimental and other empirical work, conclusions are drawn from data via theoretically supported inference.  (See \citet{longino1983body} for more on this point.)  In this case, we draw (conservative) inferences based on a combination of modeling and historical work.}

What is perhaps most interesting about the results we have presented is not that they show what can work, but rather the insight they provide into how those strategies work.  As we have emphasized above, in our discussion on unwitting propagandists, the basic mechanism at play in our models is that the propagandist biases the total evidence on which the policy-makers update their beliefs.  The fact that each result that is shared is, in some sense, ``real'' ends up being irrelevant, because it is the statistical properties of the total available evidence that matter.  For this reason, we see that producing results favorable to industry---say, through industry funding---is not necessary to manufacture favorable public beliefs, at least in cases where the evidence is probabilistic and spurious results are possible.

On this point: one might have expected that actually producing biased science would have a stronger influence on public opinion than merely sharing others' results.  But when one compares the two treatments we consider above, there are strong senses in which the less invasive, more subtle strategy of selective sharing is \emph{more} effective than biased production, all things considered, particularly when the scientific community is large and the problem at hand is difficult (or the power of generic experiments is low).  The reason is that such a scientific community will produce, on its own, plenty of research that, taken without proper context, lends credence to falsehoods.  Merely sharing this already-available evidence is cost-effective, and much less risky than producing one's own research---which, after all,  can cost a great deal to produce, fail to generate the desired results, and ultimately be disqualified or ignored because of the association with industry.  From this point of view, producing industry science (or even outright fraud) is simply not worth it.  In many cases, another more effective, less expensive, and more difficult to detect strategy is available.

Indeed, if industrial interests do wish to spend their money on science, perhaps rather than producing their own results and publishing them selectively, they would do better to adopt a different strategy, which \citet{holman2017experimentation} have called ``industrial selection''.  The idea behind industrial selection is that there are many experimental protocols, methodologies, and even research questions that scientists may adopt, often for their own reasons.  Some protocols may tend to produce more industry-friendly results than others.\footnote{\citet{holman2017experimentation} consider, for instance, the case of a class of drugs intended to stop heart failure (\citet{MooreDM}).  In that case, studies that used reducing arrhythmia as a proxy for preventing heart failure found that the drugs were highly successful, whereas studies that directly measured death rates from heart failure later showed that the drugs were substantially increasing the likelihood of death.  Industry swayed scientific practice by funding the former type of study, but not the latter.}  Industrial selection involves identifying methods already present among the community of scientists that tend to favor one's preferred outcome, and then funding scientists who already use those methods.  This extra funding works to amplify these scientists' results, by allowing them to publish more, perform higher powered studies, and train more students who will go on to produce consonant results.  Indeed, there is a strong potential feedback effect between industrial selection and selective sharing: by increasing the number of ambient industry-favorable results in a scientific literature, and then further amplifying those results by sharing them selectively, propagandists can have an even stronger effect.

It is worth emphasizing that industrial selection does not require some scientists to have adopted bad or unreliable research methods; it may also just be that they have focused on research topics that promote an industrial agenda.  For instance, \citet[p. 17]{Oreskes+Conway} describe how the Tobacco Industry Research Council supported the work of noted asbestos researcher Wilhelm C. Hueper of the National Cancer Institute.  Hueper's research was not methodologically flawed; to the contrary, it very convincingly established the (real) link between lung cancer and asbestos.  But for the tobacco industry, it was valuable to have evidence that \emph{other} environmental factors played a role in lung cancer, since that could be used to obfuscate the role their own product played.  Taken together, then, selective sharing and industrial selection suggest those interested in dampening industry influence should turn their attention to the more subtle ways propagandists can come to bias epistemic communities.  This will to a certain extent require philosophers of science to reorient their focus, as extant work tends to address explicit, direct ways industry can impact inquiry.

It is natural to ask, given the past success of the Tobacco Strategy and the results we have described, whether there are ways in which policy makers or members of the public can protect themselves against a propagandist using the methods described.  Of course, one natural suggestion is to stop listening to propagandists.  But this is difficult to do, in large part because propagandists pose as (or, in some cases, \emph{are}) credentialed scientists.\footnote{Indeed, perhaps the most striking message of \citet{Oreskes+Conway} is that the most effective propagandists have been prestigious scientists, often coming from different fields, with a political ax to grind.}  Another possibility, perhaps more realistic, is to gather evidence from as many scientists as possible: as we discuss above, policy makers connected to more scientists tend to do better, because the propagandist has to produce more evidence to effectively influence such policy makers' beliefs.\footnote{We note that connecting to more scientists was particularly effective in the selective sharing treatment when ``double-counting'' was eliminated, suggesting that policy makers would do well to attend to cases where a small number of studies are used repeatedly to prove a point, when other studies are also available.}  Connecting to many scientists may also allow policy makers to  to identify propagandists on statistical grounds: indeed, as \citet{holman2015problem} point out, in some cases it may be possible to identify biased agents by comparing the statistics of their results with those of other agents in their network.  Finally, and along similar lines, it might also be possible, at least in some circumstances, for policy makers to adjust their updating rule to accommodate the possibility that some of the evidence they receive is exaggerated, filtered, or otherwise intended to mislead.\footnote{We are grateful to David \citet{Manheim} for emphasizing this latter possibility, which is also quite close to what \citet{holman2015problem} propose.}  On the other hand, as \citet{OConnor+WeatherallPolarization} show, some apparently reasonable heuristics for updating belief in light of unreliable sources can lead to bad epistemic outcomes, including stable polarization.  We think that it would be valuable, in future work, to study how these latter two reactions might be implemented, and how and when they would be effective.

We will conclude by noting a few ways in which the incentive structure of science contributes to the success of the would-be propagandist, particularly when employing the selective sharing strategy.  As we noted above, selective sharing is most effective when (a) the problem is difficult (in the sense that $p_B$ is close to $.5$); (b) the number of data points per scientist per round is small; (c) there are many scientists working on a given problem; and (d) policy makers are connected to few scientists.  This situation interacts with how scientists are rewarded for their work in complex, troubling ways.\footnote{Previous authors have assumed that scientists are motivated by a quest for credit, in the same way normal people seek money or goods.  `Credit economy' models have been used, for example, to show why scientists might divide labor (\citet{kitcher1990division}) or commit fraud (\citet{bright2017fraud}).}  For instance, although there are some incentives for scientists to make their work accessible and to try to communicate it with the public and policy makers, it continues to be the case that academic scientists are evaluated on the basis of their research productivity---and there are sometimes reputational hazards associated with ``popularizing'' too much.  This sort of effect can limit the number of scientists to whom policy makers are connected.

But a more troubling way in which propaganda and incentives intersect concerns the role of low-powered studies.  Scientists are strongly encouraged to publish research articles, and to do so often.  Moreover, publishing positive results is strongly preferred to null results.  Finally, scientists work with limited funds that they must use to meet many demands.  All of these considerations lead to strong incentives to run studies with relatively low power (particularly, fewer data points) whenever possible (\citet{smaldino_natural_2016}).\footnote{The ubiquity of low-powered studies in some fields, such as neuroscience (\citet{button_power_2013}; \citet{szucs_empirical_2017}), ecology (\citet{lemoine_underappreciated_2016}), and evolutionary biology (\citet{hersch_power_2004}), for instance, has been remarked upon and widely discussed within those fields. \citet{smaldino_natural_2016} perform a meta-analysis of work in the behavioral sciences over the last 60 years and argue that despite repeated calls to increase power, little change has occurred.  In addition, they show why scientific communities might move towards producing low quality, low effort studies.}  Of course, most scientific fields recognize that, all else being equal, higher powered studies are more rigorous, more reliable, and make for all-around better science.  But there is little incentive, and often few resources, for individual scientists to produce consistently high quality studies.  And as we have seen, low powered studies are more likely to spuriously support erroneous conclusions. In cases where industry is not involved, this is not as worrying: spurious results can be detected via meta-analysis. But in the presence of a propagandist, they become a powerful tool in shaping public belief.

Discussions of low-powered science often focus on their role in controversies about particular results, in the context of the so-called ``replication crisis'' in the behavioral sciences (and especially psychology) (\citet{collaboration_estimating_2015}).  And the regular publication and acceptance of ultimately incorrect results is certainly worrying.  But the arguments and results we have given here suggest that even more is at stake: the same sorts of conditions that led to a replication crisis in the first place also provide extra fodder for industrial propaganda.  This makes reshaping scientists' incentives to publish many, low-powered ``least-publishable units'' crucial to foiling propagandists. For instance, scientists should be granted more credit for statistically stronger results---even in cases where they happen to be null.  Scientists should not be judged on the number of publications they produce, but on the quality of the research in those articles, so that a single publication with a very high powered study, even if it shows a null result, should be ``worth'' more than many articles showing surprising and sexy results, but with low power.  Similar reasoning suggests that, given some fixed financial resources, funding bodies should allocate those resources to a few very high-powered studies, rather than splitting it up into many small grants with only enough funding to run lower-powered studies. Or a more democratic alternative might involve scientists who work together to generate and combine smaller studies before publishing, making it more difficult for the propagandist to break off individual, spurious results.

\appendix
\section{Semi-analytic results}

Here we describe a semi-analytic treatment of the expected progress of the models presented in section \ref{sec:selectiveSharing}.

Consider a population of $K$ scientists each performing $n$ trials each round, solving a problem with $p_B = .5 + \epsilon$.  Consider a policy maker connected to $k$ scientists, with credence $B_t$ at time $t$ in the true proposition that arm B is advantageous.  This observer will update her beliefs by conditionalization in light of two sources of evidence, from the scientists she is connected to and from the propagandist.  Thus, at the end of the round, we will have:
\begin{equation}
B_{(t+1)} = Y_P\times Y_S\times B_t,
\label{eq:gen}
\end{equation}
where $Y_S$ represents the contribution from scientists and $Y_p$ represents the contribution from the propagandist.  What are these contributions on average?

In a generic round of play, the average result for each scientist performing the informative action should be $E=n*p_B$, i.e., the expectation value for $n$ draws from a binomial distribution with probability $p_B$.  We estimate $Y_s$ by supposing that the policy maker receives, from each scientist performing action B that they are connected to, the average (i.e., expected) result, and update.  This yields:
\begin{equation}
Y_S = \left(\frac{P(n*(.5+\epsilon)|H)}{P(n*(.5+\epsilon))}\right)^{r*k}
\label{eq:YS}
\end{equation}
where $r\leq1$ is the fraction of scientists in the population performing the informative action, so $r*k \leq k$ approximates the number of scientists connected to the observer who are performing the informative action.  Here $H$ is the policy maker's hypothesis about the success rate of arm B.  When multiplied by the current belief, $B_t$, as in equation \ref{eq:gen}, this is just Bayes rule for the expected evidence.  Notice that this equation does not account for the fact that $B$ will change as an agent updates on different sets of evidence within a time step.  We should think of this approximation as most useful for cases where $B$ is changing slowly.

One can show by direct computation that $Y_S$ reduces to:
\begin{equation}
Y_S = \left(\frac{1}{B_t+(1-B_t)X}\right)^{r*k},
\label{eq:YSdeets}
\end{equation}
where
\[X=\left(\frac{1-2\epsilon}{1+2\epsilon}\right)^{2n\epsilon}.\]
Observe that $X<1$, and so for any value of $B_t$, $Y_S>1$.  It follows that updating on evidence from the scientists will, on average, increase credence that arm B is preferable.  Moreover, one can see that as $n$ and $\epsilon$ each increase, $X$ will decrease, and thus $Y_S$ will increase.  Likewise, $Y_S$ will increase as $r$, the number of scientists testing the preferred arm,  increases, and as $k$, the number of scientists a policy maker listens to, increases. Of course, in the actual model stochastic effects can defeat this; we are describing average behavior.

Note that as $B_t$ decreases, $Y_S$ also becomes larger, so as observers become more skeptical, evidence from scientists will become more powerful.  Importantly this means that a set of parameter values alone will not determine the average behavior of the model: the belief state of a policy maker $B_t$ and the belief states of the scientists, as represented by $r$, will influence which way policy maker beliefs trend.

Now consider $Y_P$.  Here we consider all (and only) spurious results produced by the population of scientists each round.  First, observe that we can calculate a rate at which these will occur if all scientists perform the informative action, given by $z = P(x<n/2)\times K$, where $x$ is the number of times arm B pays off. This expression is just the probability of getting a spurious result---one that suggests $P_B <.5$---times the number of scientists in the population.  For instance, for even $n$, we have:
\begin{equation}
P(x<n/2) = \sum_{i=0}^{n/2-1}\frac{n!}{i!(n-i)!}(P_B)^i(1-P_B)^{n-i}
\label{eq:bindist}
\end{equation}
This is the sum of the probabilities of drawing each possible $x<n/2$ from a binomial distribution with $n$ draws and $P_B$.

We can also calculate an expectation value for such spurious draws, $\tilde{E}$, which is given by (again for even $n$):

\begin{equation}
\tilde{E} = \frac{1}{P(x<n/2)}\sum_{i=0}^{n/2-1}\frac{in!}{i!(n-i)!}(P_B)^i(1-P_B)^{n-i}
\label{eq:spuriousexp}
\end{equation}

This equation multiplies the value of each possible spurious draw, $i$, by the probability of that draw, conditional on the assumption that the draw will be spurious.

This leads to an estimate for $Y_P$ of:
\begin{equation}
Y_P = \left(\frac{1}{B_t+(1-B_t)\tilde{X}}\right)^{r*z},
\end{equation}
where now
\[\tilde{X}=\left(\frac{1-2\epsilon}{1+2\epsilon}\right)^{2\tilde{E}-n}\]
Once again, $r$ is the fraction of the population of scientists performing the informative action, while $z$, remember, is the rate of spurious results.  Observe that, by definition $\tilde{E}<n/2$, and so whatever else is the case $(2\tilde{E}-n) <1$, so $\tilde{X} > 1$, which means $Y_P$ is always less than 1.  Hence the propagandist will always tend to decrease the observer's confidence in the true proposition, and this effect will be stronger when more scientists take the informative action and the rate of spurious results is higher (more on this shortly).

Beyond these observations, it is harder to analyze the dependence of $Y_P$ on other parameters, on account of the complexity of the function $\tilde{E}$.  An analytic simplification of the expression for $\tilde{E}$ is apparently not available, but a numerical analysis reveals that for small $\epsilon$, $\tilde{E}$ is very well approximated for sufficiently large $n$ by a linear function in $n$, of the form $\tilde{E}=-.62 + 0.47 n$.  (This expression is determined by taking a linear best fit in Mathematica.)   Thus we have:
\begin{equation}
\tilde{X}=\left(\frac{1-2\epsilon}{1+2\epsilon}\right)^{-.06n-1.24}
\end{equation}
We see that as $\epsilon$ becomes larger, $\tilde{X}$ becomes smaller (i.e., approaches 1), and thus $Y_P$ approaches 1, so that the propagandist is less effective.  Notice though that as $n$ increases in this equation, $\tilde{X}$ becomes larger, thus $Y_P$ gets smaller, so the propagandist gets \emph{more} effective.  Thus $n$ influences $Y_P$ in two, countervailing ways.  A large $n$ makes $z$, the probability of a spurious result, smaller.  On the other hand, when spurious results do occur, they are more convincing for large $n$.  In addition, $\tilde{X}$ tends to grow slowly in $n$ and whether the propagandist ends up swaying the observer depends on the relative scaling of $Y_P$ and $Y_S$ with $n$, which in turn depends on $\epsilon$.  We also have that, as $B_t$ approaches 1, $Y_P$ approaches 1, so that the propagandist becomes less effective.  (Thus, both the propagandist and the scientists are more effective when policy makers have low credence in arm B.)

A final consideration concerns $z$, the rate of spurious results.  Clearly as $z$ increases, the propagandist becomes more effective.  This will happen as $K$, the number of scientists, increases.  (Though if policy makers listen to all the scientists, $k$ will increase as $K$ does, strengthening $Y_S$ as well.) The rate $z$ will also increase as $r$, the fraction of scientists performing the informative action, increases, though an increase in $r$ will increase both $Y_S$ and $Y_P$.  There is also a dependence in $z$ on $P(x<n/2)$.  It is difficult to get a good first order approximation for this function, but by studying plots one can characterize some qualitative features.  In general, for fixed $n$, increasing $\epsilon$ decreases $P(x<n/2)$, and thus decreases $z$, making the propagandist less successful.

For sufficiently large $n$, we find that as $n$ increases, $P(x<n/2)$  decreases, again making the propagandist less effective.  This is what one would expect: the probability of spurious results decreases as $n$ increases. But what is surprising is that for small $n$, there is a regime in which increasing $n$ actually increases $P(x<n/2)$.  For sufficiently large $\epsilon$ (say $>.1$ or so), this regime is a small part of $n$-space.  But for small $\epsilon$ (say, $\epsilon\approx .001$), $P(x<n/2)$ actually increases with $n$ for all values of $n$ that we study.  So increasing $n$ under these circumstances makes the propagandist more effective.  Again, an increase in $n$ also makes scientists more effective, so the overall influence of $n$ will depend on whether it increases $Y_P$ or $Y_S$ to a greater degree.

\section*{Acknowledgments}

This material is based upon work supported by the National Science Foundation under grant number 1535139 (O'Connor).  The majority of the work on this project was completed while O'Connor and Weatherall were Visiting Fellows at the Australian National University; we are grateful to the Research School for Social Science and the School of Philosophy for support.  We are also grateful to audiences at the ANU, University of Groningen, UC Irvine, UC Merced, UC San Diego, and the 2018 PPE Society Meeting for helpful feedback on talks related to this work, and to Jeff Barrett, Craig Callender, Bennett Holman, David Manheim, Aydin Mohseni, Mike Schneider, and Chris Smeenk for helpful discussions.


\begin{thebibliography}{35}
\expandafter\ifx\csname natexlab\endcsname\relax\def\natexlab#1{#1}\fi
\expandafter\ifx\csname url\endcsname\relax
  \def\url#1{{\tt #1}}\fi
\expandafter\ifx\csname urlprefix\endcsname\relax\def\urlprefix{URL }\fi

\bibitem[{Bala and Goyal(1998)}]{venkatesh1998learning}
Bala, V.  and Goyal, S. [1998]:  `Learning from neighbors',
\newblock {\em Review of Economic Studies\/}, {\textbf{ 65}\/}(3), pp.
  595--621.

\bibitem[{Bright(2017)}]{bright2017fraud}
Bright, L.~K. [2017]:  `On fraud',
\newblock {\em Philosophical Studies\/}, {\textbf{ 174}\/}(2), pp. 291--310.

\bibitem[{Button \emph{et~al.}(2013)Button, Ioannidis, Mokrysz, Nosek, Flint,
  Robinson, and Munafò}]{button_power_2013}
Button, K.~S., Ioannidis, J. P.~A., Mokrysz, C., Nosek, B.~A., Flint, J.,
  Robinson, E. S.~J.  and Munafò, M.~R. [2013]:  `Power failure: why small
  sample size undermines the reliability of neuroscience',
\newblock {\em Nature Reviews Neuroscience\/}, {\textbf{ 14}\/}(5), pp.
  365--376.

\bibitem[{Douglas(2009)}]{DouglasSPVFI}
Douglas, H. [2009]: {\em Science, policy, and the value-free ideal\/},
\newblock Pittsburgh, PA: University of Pittsburgh Press.

\bibitem[{Elliott(2011)}]{ElliottLPGFY}
Elliott, K.~C. [2011]: {\em Is a Little Pollution Good For You?\/},
\newblock Oxford, UK: Oxford University Press.

\bibitem[{Frey and {\v{S}}e{\v{s}}elja(2018)}]{frey2018epistemic}
Frey, D.  and {\v{S}}e{\v{s}}elja, D. [2018]:  `What Is the Epistemic Function
  of Highly Idealized Agent-Based Models of Scientific Inquiry?',
\newblock {\em Philosophy of the Social Sciences\/}.

\bibitem[{Hersch and Phillips(2004)}]{hersch_power_2004}
Hersch, E.~I.  and Phillips, P.~C. [2004]:  `Power and potential bias in field
  studies of natural selection',
\newblock {\em Evolution; International Journal of Organic Evolution\/},
  {\textbf{ 58}\/}(3), pp. 479--485.

\bibitem[{Holman and Bruner(2017)}]{holman2017experimentation}
Holman, B.  and Bruner, J. [2017]:  `Experimentation by industrial selection',
\newblock Forthcoming in Philosophy of Science.

\bibitem[{Holman and Bruner(2015)}]{holman2015problem}
Holman, B.  and Bruner, J.~P. [2015]:  `The problem of intransigently biased
  agents',
\newblock {\em Philosophy of Science\/}, {\textbf{ 82}\/}(5), pp. 956--968.

\bibitem[{Ioannidis(2005)}]{ioannidis2005most}
Ioannidis, J.~P. [2005]:  `Why most published research findings are false',
\newblock {\em PLoS medicine\/}, {\textbf{ 2}\/}(8), pp. e124.

\bibitem[{Kitcher(1990)}]{kitcher1990division}
Kitcher, P. [1990]:  `The division of cognitive labor',
\newblock {\em The journal of philosophy\/}, {\textbf{ 87}\/}(1), pp. 5--22.

\bibitem[{Kitcher(2003)}]{KitcherSTD}
Kitcher, P. [2003]: {\em Science, truth, and democracy\/},
\newblock New York, NY: Oxford University Press.

\bibitem[{Kitcher(2011)}]{KitcherSDS}
Kitcher, P. [2011]: {\em Science in a democratic society\/},
\newblock Amherst, NY: Prometheus Books.

\bibitem[{Lemoine \emph{et~al.}(2016)Lemoine, Hoffman, Felton, Baur, Chaves,
  Gray, Yu, and Smith}]{lemoine_underappreciated_2016}
Lemoine, N.~P., Hoffman, A., Felton, A.~J., Baur, L., Chaves, F., Gray, J., Yu,
  Q.  and Smith, M.~D. [2016]:  `Underappreciated problems of low replication
  in ecological field studies',
\newblock {\em Ecology\/}, {\textbf{ 97}\/}(10), pp. 2554--2561.

\bibitem[{Longino and Doell(1983)}]{longino1983body}
Longino, H.  and Doell, R. [1983]:  `Body, bias, and behavior: A comparative
  analysis of reasoning in two areas of biological science',
\newblock {\em Signs: Journal of Women in Culture and Society\/}, {\textbf{
  9}\/}(2), pp. 206--227.

\bibitem[{Longino(1990)}]{LonginoSSK}
Longino, H.~E. [1990]: {\em Science as social knowledge: Values and objectivity
  in scientific inquiry\/},
\newblock Princeton, NJ: Princeton University Press.

\bibitem[{Longino(2002)}]{LonginoFK}
Longino, H.~E. [2002]: {\em The fate of knowledge\/},
\newblock Princeton, NJ: Princeton University Press.

\bibitem[{Manheim(2018)}]{Manheim}
Manheim, D. [2018]:  `Comment on {``How to Beat Science and Influence
  People''}',
\newblock
  Https://medium.com/@davidmanheim/comment-on-how-to-beat-science-and-influence-people-https-arxiv-org-abs-1801-01239-5e96f0ad49bd.

\bibitem[{Michaels(2008)}]{MichaelsDISP}
Michaels, D. [2008]: {\em Doubt is Their Product\/},
\newblock Oxford, UK: Oxford University Press.

\bibitem[{Michaels and Monforton(2005)}]{michaels2005manufacturing}
Michaels, D.  and Monforton, C. [2005]:  `Manufacturing uncertainty: contested
  science and the protection of the public’s health and environment',
\newblock {\em American journal of public health\/}, {\textbf{ 95}\/}(S1), pp.
  S39--S48.

\bibitem[{Moore(1995)}]{MooreDM}
Moore, T.~J. [1995]: {\em Deadly medicine: why tens of thousands of heart
  patients died in America's worst drug disaster\/},
\newblock New York, NY: Simon \& Schuster.

\bibitem[{Norr(1952)}]{norr1952cancer}
Norr, R. [1952]:  `Cancer by the carton',
\newblock {\em Reader's Digest\/}, {\textbf{ 61}\/}(368), pp. 7--8.

\bibitem[{O'Connor and Weatherall(2017)}]{OConnor+WeatherallPolarization}
O'Connor, C.  and Weatherall, J.~O. [2017]:  `Scientific Polarization',
\newblock ArXiv:1712.04561 [cs.SI].

\bibitem[{O'Connor and Weatherall(2019)}]{OConnor+Weatherall}
O'Connor, C.  and Weatherall, J.~O. [2019]: {\em The Misinformation Age: How
  False Beliefs Spread\/},
\newblock New Haven, CT: Yale University Press.

\bibitem[{{Open Science Collaboration}(2015)}]{collaboration_estimating_2015}
{Open Science Collaboration} [2015]:  `Estimating the reproducibility of
  psychological science',
\newblock {\em Science\/}, {\textbf{ 349}\/}(6251), pp. aac4716.
\newline <http://science.sciencemag.org/content/349/6251/aac4716>

\bibitem[{Oreskes and Conway(2010)}]{Oreskes+Conway}
Oreskes, N.  and Conway, E.~M. [2010]: {\em Merchants of Doubt\/},
\newblock New York, NY: Bloomsbury Press.

\bibitem[{Romero(2016)}]{romero2016can}
Romero, F. [2016]:  `Can the behavioral sciences self-correct? A social
  epistemic study',
\newblock {\em Studies in History and Philosophy of Science Part A\/},
  {\textbf{ 60}\/}, pp. 55--69.

\bibitem[{Rosenstock \emph{et~al.}(2017)Rosenstock, Bruner, and
  O’Connor}]{rosenstock2017epistemic}
Rosenstock, S., Bruner, J.  and O’Connor, C. [2017]:  `In Epistemic Networks,
  Is Less Really More?',
\newblock {\em Philosophy of Science\/}, {\textbf{ 84}\/}(2), pp. 234--252.

\bibitem[{Smaldino and McElreath(2016)}]{smaldino_natural_2016}
Smaldino, P.~E.  and McElreath, R. [2016]:  `The natural selection of bad
  science',
\newblock {\em Open Science\/}, {\textbf{ 3}\/}(9), pp. 160384.

\bibitem[{Szucs and Ioannidis(2017)}]{szucs_empirical_2017}
Szucs, D.  and Ioannidis, J. P.~A. [2017]:  `Empirical assessment of published
  effect sizes and power in the recent cognitive neuroscience and psychology
  literature',
\newblock {\em PLOS Biology\/}, {\textbf{ 15}\/}(3), pp. e2000797.
\newline
  <http://journals.plos.org/plosbiology/article?id=10.1371/journal.pbio.2000797>

\bibitem[{Weatherall and O'Connor(2018)}]{OConnor+WeatherallConformity}
Weatherall, J.~O.  and O'Connor, C. [2018]:  `Conformity in Scientific
  Networks',
\newblock ArXiv:1803.09905 [physics.soc-ph].

\bibitem[{Woodward(2003)}]{woodward2005making}
Woodward, J. [2003]: {\em Making things happen: A theory of causal
  explanation\/},
\newblock Oxford university press.

\bibitem[{Wynder \emph{et~al.}(1953)Wynder, Graham, and
  Croninger}]{wynder1953experimental}
Wynder, E.~L., Graham, E.~A.  and Croninger, A.~B. [1953]:  `Experimental
  production of carcinoma with cigarette tar',
\newblock {\em Cancer Research\/}, {\textbf{ 13}\/}(12), pp. 855--864.

\bibitem[{Zollman(2007)}]{zollman2007communication}
Zollman, K.~J. [2007]:  `The communication structure of epistemic communities',
\newblock {\em Philosophy of science\/}, {\textbf{ 74}\/}(5), pp. 574--587.

\bibitem[{Zollman(2010)}]{zollman2010epistemic}
Zollman, K.~J. [2010]:  `The epistemic benefit of transient diversity',
\newblock {\em Erkenntnis\/}, {\textbf{ 72}\/}(1), pp. 17.

\end{thebibliography}
\end{document}